\documentclass[aps,prc,showpacs]{revtex4}
\usepackage{graphicx}
\usepackage[dvips]{color}
\usepackage{colordvi}
\begin{document}
\title{Diagrammatic calculation of thermodynamical quantities in nuclear matter}\thanks{Research supported in part by: the Polish State 
Committee for Scientific Research, grant No. 2 P03B 059 25}
\author{     V. Som\`a\footnote{Electronic address~:
vittorio.soma@ifj.edu.pl}}
\affiliation{
Institute of Nuclear Physics, PL-31-342 Cracow, Poland}
\author{     P. Bo\.{z}ek\footnote{Electronic address~:
piotr.bozek@ifj.edu.pl}}
\affiliation{
Institute of Nuclear Physics, PL-31-342 Cracow, Poland\\
and\\
Institute of Physics, Rzeszow University, PL-35-959 Rzeszow, Poland}

\date{\today}

\begin{abstract}
In medium $T-$matrix calculations for symmetric nuclear matter at zero and
finite temperatures are presented. The internal energy is calculated from the 
Galitskii-Koltun's sum rule and from the summation of the diagrams for the 
 interaction energy. 
 The pressure at finite temperature is obtained
from the generating functional form of the thermodynamic potential.
The entropy at high temperature is estimated and compared to expressions
corresponding to a quasiparticle gas.
\end{abstract}

\pacs
{\bf  21.30.Fe, 21.65.+f, 24.10.Cn}

\maketitle

\section{Introduction}

\label{intro}

The description of nuclear matter and its thermodynamic properties is
an important issue for the modeling of hot neutron stars and 
intermediate energy heavy-ion collisions. A possible approach consists in
trying
to calculate the bulk properties of the many-body system, starting from a
free N-N potential.
Strong N-N interactions induce short range correlations 
in nuclear matter, which have to be treated consistently in the dense system.
The effect of the short range interactions on the binding
energy depends on the particular free N-N potential used; moreover it is known
that, in order to reproduce the empirical saturation point in symmetric nuclear
matter, three-body forces must be considered. 
In the present paper we present a study restricted to a model
using the two-body CD-Bonn potential only, while three-body interactions 
will be included in a further publication. 

A thermodynamically consistent approximation which resums the short range
correlations can be constructed
from a suitably chosen generating functional \cite{kadBaym,baym}.
For nuclear interactions the generating functional must at least include 
the sum of ladder type diagrams, this choice leads to the in medium
$T-$matrix
approximation \cite{kadBaym}. This is the approximation scheme adopted in our
study.
At zero temperature it yields results for binding energy, pressure, and
single particle-energy compatible with each other
 \cite{Bozek:2001tz,Bozek:2002mw}.
Thermodynamical relations are fulfilled, including the celebrated
Hugenholz-Van Hove  and Luttinger identities \cite{hvh,luttinger2}.
The thermodynamically consistent in-medium $T-$matrix
approach at zero temperature
\cite{Bozek:2002tz,Bozek:2002ry,Dewulf:2002gi,Dewulf:2003nj}
yields  results for the binding energy similar to extensive variational and
Brueckner-Hartree-Fock approaches \cite{vcs2,vcs1,baldonstar,Baldo:2001mw}.

Much less studies are available at finite temperatures.
Some nuclear matter calculations at finite temperature within the
Brueckner-Hartree-Fock approach have been performed
\cite{Baldo:Ferreira,Zuo:2006wx}.
The in medium $T-$matrix scheme based on
finite-temperatures Green's functions can be easily applied to calculate
the internal energy or the single-particle properties
\cite{Bozek:1998su,Bozek:2002em,Frick:2003sd,Frick2,Frick:2004th},
however no estimates for the pressure or the entropy of the interacting 
nucleon system with short range correlations are available within this
approach.
At zero temperature the pressure $P$ can be 
calculated from the binding energy per particle $E/N$ 
using the thermodynamic relations 
$P=\rho(\mu-E/N)$ or $P=\rho^2\frac{\partial(E/N)}{\partial \rho}$, where 
$\rho$ is the nuclear matter density and $\mu$ is the chemical potential
\cite{Bozek:2001tz}. In principle, the thermodynamic identities
\begin{equation}
\left(\frac{\partial PV}{\partial T}\right)_{\mu   ,  V}
=\frac{E+PV-\mu N}{T}  \ \ ;  \ \ 
T\left(\frac{\partial S}{\partial T}\right)_{\mu
  ,  V}=\left(\frac{\partial( E-\mu N)}{\partial T}\right)_{\mu   ,  V}
\end{equation}
could be  integrated in order to obtain the pressure or the entropy $S$
 at non-zero temperatures ($V$ is the volume of the system).
In practice, the above equations cannot be employed to estimate numerically 
the pressure since an inaccurate evaluation of the numerators
 yields uncontrollable errors for the entropy at small temperatures. 
The pressure can be  calculated instead from 
the diagrammatic expansion of 
the thermodynamic potential. 
Through the relation
\begin{equation}
TS=E+PV-\mu N \ \: 
\end{equation}
the entropy 
can be reliably estimated for sufficiently large temperatures.
In section \ref{sec:gk} we discuss diagrammatic procedures for calculating the
energy of the system and compare the result to the Galitskii-Koltun's formula.
By varying the strength of the interaction 
potential in the  diagrammatic formula for the internal
energy  we obtain an expression
for the pressure of the interacting system (section \ref {sec:ps}).
 The entropy is also estimated and compared 
with a reduced formula and  with the  result for a free gas of Fermions with
in-medium modified masses.

\section{Energy of the interacting system}

\label{sec:gk}

The (total) internal  energy per particle can be calculated as the expectation
value of the Hamiltonian of the system $H = H_{kin} + H_{pot}$
\begin{equation}
\label{eq:be}
\frac{E}{N} = \frac{1}{\rho} 
\left [
\frac{\langle H_{kin} \rangle}{V} + \frac{\langle H_{pot} \rangle}{V} 
\right ]\: .
\end{equation}
When we evaluate the right hand side of (\ref{eq:be}) in momentum space
we see that the kinetic part is 
\begin{equation}
\label{eq:kin}
\langle H_{kin} \rangle = V \int \frac{d^3{p}}{(2\pi)^3} 
\frac{d \omega}{2\pi}
\frac{\mathbf{p}^2}{2m} A(\mathbf{p},\omega) f(\omega) \: ,
\end{equation}
while the potential term  takes the form
\begin{equation}
\label{eq:pot}
\langle H_{pot} \rangle = \displaystyle \frac{V}{2} 
\int \frac{d^3P}{(2\pi)^3} \frac{d^3{k}}{(2\pi)^3} 
\frac{d^3{k'}}{(2\pi)^3} 
\frac{d\Omega}{2\pi} V(\mathbf{k},\mathbf{k'})
\langle\mathbf{k'}|G_2^<(\mathbf{P},\Omega)|\mathbf{k}\rangle \ .
\end{equation}
Here $A(\mathbf{p},\omega)$ is the spectral function, $f(\omega)$ is the Fermi 
distribution and $V(\mathbf{k},\mathbf{k'})$ the interaction potential
(we skip the spin and isospin indices in the notation).
$\langle\mathbf{k'}|G_2^<(\mathbf{P},\Omega)|\mathbf{k}\rangle$ is the two
particle Green's function with a common  time for the incoming lines and a
common time for the 
 outgoing lines. The outgoing time is set to be larger than the incoming 
time (this corresponds to $^{<}$~ordering on the 
real-time contour)  \cite{keldysh,Danielewicz:1982kk};
after integration over the total energy of the pair
$\Omega$,  the incoming and outgoing times are set to the same value.
The presence of the full two-particle propagator requires the use of 
diagrammatic calculation techniques, in particular the two-particle 
Green's function must be calculated within the chosen approximation scheme.
Alternatively, it is possible to determine the total energy in a simpler way 
through the Galitskii-Koltun's sum rule \cite{galitskii,martinschwinger,koltun}
\begin{equation}
\label{gksumrule}
\frac{E}{N} = \frac{1}{\rho}
\int \frac{d^3{p}}{(2 \pi)^3} \frac{d \omega}{2 \pi}
\left [ \frac{\mathbf{p}^2}{2m} + \omega \right ] 
A(\mathbf{p},\omega) f(\omega )\: .
\end{equation}
For conserving approximations, as well as in case when the full (exact)
solutions for the two-particle Green's
function and the spectral function are used,
the expression for the energy of the
 system in the form of the above sum rule is 
equivalent to the direct calculation (\ref{eq:be}).
On the other hand this does not remain valid when three-body forces are
included: in that case the Galitskii-Koltun's sum rule cannot be employed
and the energy has to be evaluated directly from the expectation value of
the Hamiltonian.

Let us first note that the simplest way of approximating a two-particle
Green's function consists in constructing a product of two single 
one-body propagators $G$.
We shall call this simple approximation \textit{non correlated} 
two-particle Green's function and indicate it with $G_2^{nc}$
(we restrict our formulas to the case when the times
 of the two incoming as well as
the two outgoing  lines are the
same):
\begin{eqnarray}
\label{g2nc}
& \langle\mathbf{k'}|{G_2^{nc \ <(>)}}(\mathbf{P},\Omega)|\mathbf{k}\rangle =
\nonumber \\
& = \displaystyle
i (2\pi)^3 \delta (\mathbf{k}-\mathbf{k'}) \int \frac{d\omega'}{2\pi}
G^{<(>)}(\mathbf{P}/2+\mathbf{k},\Omega-\omega')
G^{<(>)}(\mathbf{P}/2-\mathbf{k},\omega') \: .
\end{eqnarray}
The single-particle Green's function $G$ denotes the 
in-medium dressed propagator, which includes the self-energy 
resummed in the chosen approximation. In the 
self-consistent $T-$matrix approximation the dressed propagator involves 
a nontrivial dispersive self-energy which leads to a broad spectral function
\cite{Bozek:2002em}. 
We then introduce the in-medium two-particle scattering matrix $T$, defined by 
\begin{eqnarray}
& \langle\mathbf{k}|T^{R(A)}(\mathbf{P},\Omega)|\mathbf{k'}\rangle = 
V(\mathbf{k},\mathbf{k'})
\nonumber \\
& + \displaystyle \int \frac{d^3{p}}{(2\pi)^3} 
\frac{d^3{q}}{(2\pi)^3}
V(\mathbf{k},\mathbf{p}) 
\langle\mathbf{p}|{G_2^{nc\ R(A)}}(\mathbf{P},\Omega)|\mathbf{q}\rangle
\langle\mathbf{q}|T^{R(A)}(\mathbf{P},\Omega)|\mathbf{k'}\rangle \: ,
\end{eqnarray}
where we have used the non correlated two-particle Green's function defined
above, with the time ordering in the times of the incoming and outgoing lines
 changed to the retarded $G_2^R$ or advanced $G_2^A$ one 
(the same time ordering is chosen for the $T-$matrix)
 \cite{Danielewicz:1982kk}.
The self-consistent $T-$matrix approach is an iterative scheme involving
the calculation of the $T-$matrix, the calculation of the self-energy
\cite{Bozek:1998su}
\begin{eqnarray}
\mbox{Im}\Sigma(\mathbf{p},\omega)&=&\int   \frac{d^3{k}}{(2\pi)^3}
\frac{d{\omega^{'}}}{2\pi}\left[ \mbox{Im} 
\langle(\mathbf{p-k})/2|T^{R}(\mathbf{p+k},\Omega)|(\mathbf{p-k})/2\rangle
- \right. \nonumber \\ & &\left. 
\mbox{Im} \langle(\mathbf{p-k})/2|T^{R}(\mathbf{p+k},\Omega)|
(\mathbf{k-p})/2\rangle
\right]\left[ b(\omega+\omega^{'})+f(\omega^{'})\right]A(\mathbf{k},\omega^{'})
\end{eqnarray}
and of the Dyson equation
\begin{equation}
G^{R(A)\ -1}(\mathbf{p},\omega)=\omega-\frac{p^2}{2m}-
\Sigma^{R(A)}(\mathbf{p},\omega) \ .
\end{equation}
The  $T-$matrix approximation can be as well obtained from a 
generating functional $\Phi[G,V]$ which is a set of two-particle irreducible
diagrams \cite{baym} obtained by closing    two-particle    ladder
diagrams  with a combinatorial factor $1/2n$, where $n$ is the number of
interaction lines in the diagram. 
\begin{figure}[h]
\begin{center}
\includegraphics[width=10cm]{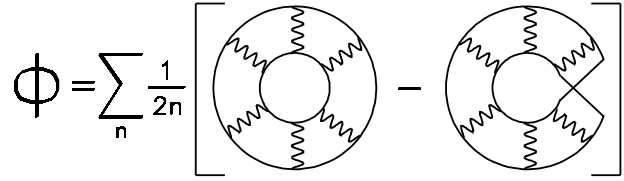}
\caption{Two-particle irreducible functional for the $T-$matrix approximation.
The sum runs over the number of interaction lines $n$ in the two-particle 
ladders.}
\label{fig:tphi}
\end{center}
\end{figure}
 The functional $\Phi$ depends on the dressed
propagators (lines in Fig. \ref{fig:tphi})
 and on the two-particle potential (wavy lines 
  in Fig. \ref{fig:tphi}).
The self-energy is then given as a functional derivative 
$\displaystyle \Sigma=\frac{\delta\Phi}{\delta G}$.
  
An approximation for $G_2$, which we shall indicate with the
italic character $\mathcal{G}_{2}$, can  be written as follows
\begin{eqnarray}
\label{app}
& \langle\mathbf{k}|\mathcal{G}_{2}^{R(A)}
(\mathbf{P},\Omega)|\mathbf{k'}\rangle = 
\langle\mathbf{k}|{G_2^{nc \ R(A)}}(\mathbf{P},\Omega)|\mathbf{k'}\rangle -
\langle\mathbf{k}|{G_2^{nc \ R(A)}}(\mathbf{P},\Omega)|\mathbf{-k'}\rangle
\nonumber \\
& + \displaystyle \int \frac{d^3{p}}{(2\pi)^3} 
\frac{d^3{q}}{(2\pi)^3} \left\{
\langle\mathbf{k}|{G_2^{nc\ R(A)}}(\mathbf{P},\Omega)|\mathbf{p}\rangle -
\langle\mathbf{k}|{G_2^{nc\ R(A)}}(\mathbf{P},\Omega)|\mathbf{-p}\rangle 
\right\}
\nonumber \\
& \times \displaystyle 
\langle\mathbf{p}|T^{R(A)}(\mathbf{P},\Omega)|\mathbf{q}\rangle 
\langle\mathbf{q}|{G_2^{nc\ R(A)}}(\mathbf{P},\Omega)|\mathbf{k'}\rangle \: .
\end{eqnarray}
We insert this expression in (\ref{eq:pot}) and we get
\begin{eqnarray}
\langle H_{pot} \rangle &=&\frac{V}{2}
\int \frac{d^3{P}}{(2\pi)^3} \frac{d^3{k}}{(2\pi)^3}
\frac{d\Omega}{2\pi} \left[
\langle\mathbf{k}|T(\mathbf{P},\Omega)|\mathbf{k}\rangle \right.
\nonumber \\
& & \left. \times
\left( \langle\mathbf{k}|{G_2^{nc}}(\mathbf{P},\Omega)|\mathbf{k}\rangle
-\langle\mathbf{k}|{G_2^{nc}}(\mathbf{P},\Omega)|\mathbf{-k}\rangle 
\right) \right]^{<} \: ,
\end{eqnarray}
where the superscript $^<$ concerns  the time ordering of the
product $T\ G_2^{nc}$.
\begin{figure}[h]
\begin{center}
\includegraphics[width=10cm]{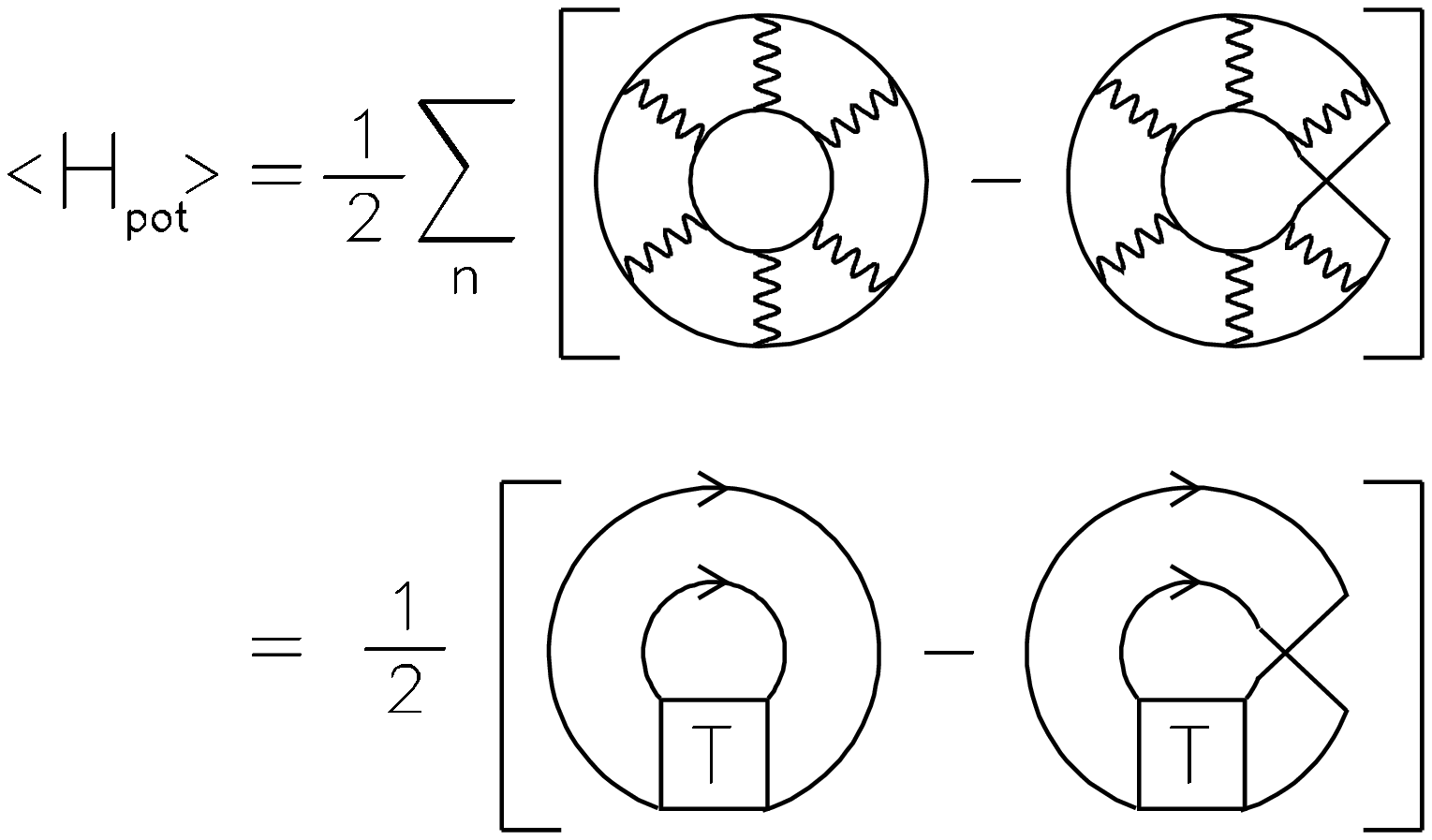}
\caption{Diagrammatic expression for the expectation value 
of the  potential energy. As in Fig. \ref{fig:tphi},
the sum runs over the number of interaction lines $n$ in the diagram.}
\label{fig:diagv}
\end{center}
\end{figure}
\begin{figure}[h]
\begin{center}
\includegraphics[width=8cm]{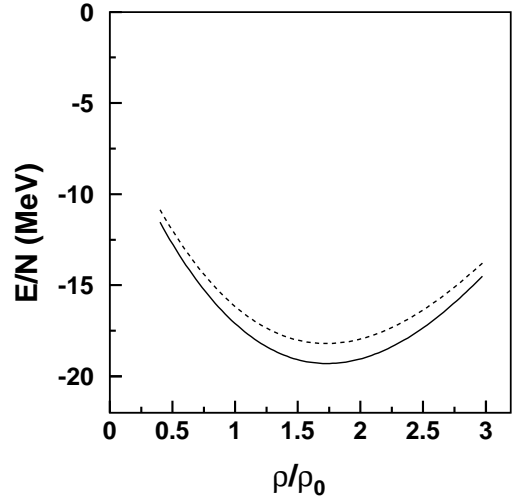}
\caption{The internal energy per particle at zero temperature 
as a function of the density (in units of the empirical
saturation density $\rho_0=0.16$ fm$^{-3}$). The solid
line represents the expectation value of the Hamiltonian (\ref{eq:be}) and 
the dashed line is the result obtained from the Galitskii-Koltun's sum rule 
(\ref{gksumrule}).}
\label{fig:energy}
\end{center}
\end{figure}
Finally, the explicit formula with  the retarded functions is
\begin{equation}
\label{eq:final}
\langle H_{pot} \rangle = \frac{V}{2} 
\int \frac{d^3{P}}{(2\pi)^3} \frac{d^3{k}}{(2\pi)^3} 
\frac{d\Omega}{2\pi}  \, b(\Omega) \, \displaystyle
\mbox{Im} \left\{ \left( \langle \mathbf{k}|T^R(\mathbf{P},\Omega)
|\mathbf{k}\rangle-
 \langle \mathbf{k}|T^R(\mathbf{P},\Omega) |\mathbf{-k}\rangle\right)\,
{G_2^{nc\ R}}(\mathbf{P},\mathbf{k},\Omega)\right\} \: ,
\end{equation}
where $b(\Omega)$ is the Bose-Einstein distribution (Fig. \ref{fig:diagv}). 
The same expression can be obtained equivalently in the
 imaginary time formalism.

We calculate the internal energy per particle
 with the use of the two methods:
\begin{enumerate}
\item from the Galitskii-Koltun's sum rule (\ref{gksumrule});
\item from diagram summation, i.e. from eq. (\ref{eq:be}) together 
with (\ref{eq:kin}) and (\ref{eq:final}).
\end{enumerate}
Results for the symmetric nuclear matter at zero temperature with the CD Bonn
potential are displayed in Fig. \ref{fig:energy}.
The calculation are performed using a numerical procedure where the energy 
range  is limited to an interval $[-\omega_c,\omega_c]$. We have tested
several values of the energy scale $\omega_c$ between
 $2\,\mbox{GeV}$ and $10\,\mbox{GeV}$, and we found 
that the Galitskii-Koltun's sum rule expression for the energy shows some
dependence on this value. On the other hand, the results obtained from the 
direct estimation of the interaction and kinetic energies are stable and
 independent on the energy range taken 
(up to an inaccuracy due to numerical discretization 
 of about $0.5$ MeV). The result from the diagram summation can be compared to 
Galitskii-Koltun's sum rule result extrapolated to infinite energy range 
(Fig. \ref{fig:energy}).
The values obtained in two ways
 differ by about $0.9$ MeV at the saturation density. 
Besides the numerical 
inaccuracies another source of the difference between
the internal energy from the Galitskii-Koltun's sum rule and from the 
expression (\ref{eq:be}) can be attributed to the effect of the
angular averaging of the two-particle propagator in the $T-$matrix
calculation \cite{Schiller:1998ff}.
Such a technical approximation is used in order to allow for a partial-wave
expansion of the in-medium $T-$matrix. 
In the following we use for the internal energy either the 
Galitskii-Koltun's sum rule result extrapolated to infinite energy range
 or the 
expression (\ref{eq:be}).
We find differences of the same order 
 using the two methods of calculating the 
internal energy also at finite temperatures (Table \ref{tab:pressure}).

As stated in section \ref{intro} the calculation using only a two-body
force are not complete, e.g. the saturation density is too large, and
only the result of a study which includes three-nucleon interactions
could be compared to an experimentally estimated equation of state of
nuclear matter. Therefore in the following we restrict ourselves only 
to the empirical saturation density $\rho_0=0.16$ fm$^{-3}$ to illustrate 
the calculation of the pressure and entropy.

\section{Pressure and entropy}

\label{sec:ps}

The pressure is related to the thermodynamical potential through
\begin{equation}
\Omega(T,\mu,V)=-PV \: .
\end{equation}
It can be shown that $\Phi$ gives a contribution to the
thermodynamical potential \cite{fw}, and in particular that
\begin{equation} 
\label{om}
\Omega=-\mbox{Tr}\{\ln[G^{-1}]\} -\mbox{Tr}\{\Sigma G\} + \Phi \: .
\end{equation}
In the real-time formalism the trace involves the integration over energy and
momenta, with the time ordering corresponding to $^<$;
however the equations are most easily derived in the imaginary time formalism,
where the trace implies a summation over Matsubara frequencies together 
with an integration
over momenta, and the Matsubara representation of the dressed single-particle 
Green's function takes the form
\begin{equation}
G(\mathbf{p},i \omega_n)=\int \frac{d\omega}{2\pi}
\frac{A(\mathbf{p},\omega)}{i\omega_n-\omega} 
\ \ .
\end{equation}
We  write the pressure as a sum of two terms
\begin{equation} 
\label{eq:pre}
P_{tot}= P_I + P_{II} \: ,
\end{equation}
where
\begin{equation} 
\label{eq:p1}
P_I = \frac{1}{V} \left [
\mbox{Tr}\{\ln[G^{-1}]\} +\mbox{Tr}\{\Sigma G\}
\right ] \: 
\end{equation}
and
\begin{equation} 
\label{eq:p2}
P_{II} = - \frac{\Phi}{V} \: .
\end{equation}
The two contributions in (\ref{eq:p1}) give respectively 
\begin{eqnarray}
\displaystyle
 \mbox{Tr}\{\ln[G^{-1}]\}&= & 
VT \displaystyle \int \frac{d^3{p}}{(2\pi)^3} \frac{d \omega}{2\pi} 
\ln (1+e^{-\beta \omega}) \left [ A(\mathbf{p},\omega) + \right.
\nonumber \\ &+ & \displaystyle
2 \, \frac{\partial\mbox{Re}\Sigma^{R}(\mathbf{p},\omega)}
{\partial\omega}\mbox{Im}G^R(\mathbf{p},\omega)
+ 2 \, \frac{\partial\mbox{Im}\Sigma^{R}(\mathbf{p},\omega)}
{\partial\omega}\mbox{Re}G^R(\mathbf{p},\omega) \left. \right ] \: 
\end{eqnarray}
and
\begin{eqnarray}
 \mbox{Tr}\{\Sigma G\} &=&
 \displaystyle VT \int \frac{d^3{p}}{(2\pi)^3} \frac{d \omega}{2\pi}
\ln (1+e^{-\beta \omega})
\left [ \right.
\frac{\partial A(\mathbf{p},\omega)}{\partial\omega}
\mbox{Re}\Sigma^R (\mathbf{p},\omega)
\nonumber \\ &
+ & \displaystyle A(\mathbf{p},\omega) 
\frac{\partial\mbox{Re}\Sigma^{R}(\mathbf{p},\omega)}{\partial\omega}
- 2 \, \frac{\partial\mbox{Im}\Sigma^{R}(\mathbf{p},\omega)}
{\partial\omega}\mbox{Re}G^R(\mathbf{p},\omega) \nonumber \\
&-& 2 \, \mbox{Im}\Sigma^{R}(\mathbf{p},\omega)
\frac{\partial\mbox{Re}G^{R}(\mathbf{p},\omega)}{\partial\omega}
\left. \right ] \: ,
\end{eqnarray}
so that 
\begin{eqnarray}
 \displaystyle
P_I & =& 
T \int \frac{d^3p}{(2\pi)^3} \frac{d \omega}{2\pi}
\ln (1+e^{-\beta \omega}) 
\left [ \right.
A(\mathbf{p},\omega)+
\nonumber \\
& + &\displaystyle
\frac{\partial A(\mathbf{p},\omega)}{\partial\omega}
\mbox{Re}\Sigma^R (\mathbf{p},\omega)
- 2 \mbox{Im}\Sigma^{R} (\mathbf{p},\omega) 
\frac{\partial\mbox{Re}G^{R}(\mathbf{p},\omega)}{\partial\omega}
\left. \right ]
\nonumber \\
&=& T \int \frac{d^3p}{(2\pi)^3} \frac{d \omega}{2\pi}
\ln (1+e^{-\beta \omega}) 
B(\mathbf{p},\omega) \: .
\end{eqnarray}
We introduce a new weight function $B(\mathbf{p},\omega)$ 
\cite{Weinhold:1997ig}; it is
 normalized similarly as the spectral function $A(\mathbf{p},\omega)$
\begin{equation}
\label{eq:intb}
\int \frac{d\omega}{2\pi} B(\mathbf{p},\omega)=1 
\end{equation}
but assumes negative as well as positive values.
The spectral function at small temperature exhibits a sharp peak
at momenta close to the Fermi momentum, causing difficulties in 
numerical calculations. In  ref. \cite{Bozek:2002em} this
problem was avoided by separating the spectral function into a smooth
background part and an approximate $\delta$ function
\begin{equation}
A(\mathbf{p},\omega)=A_{bg}(\mathbf{p},\omega)+W_p \, 2\pi \,
\delta(\omega-\omega_p) \ .
\end{equation}
\begin{figure}[h]
\begin{center}
\includegraphics[width=8cm]{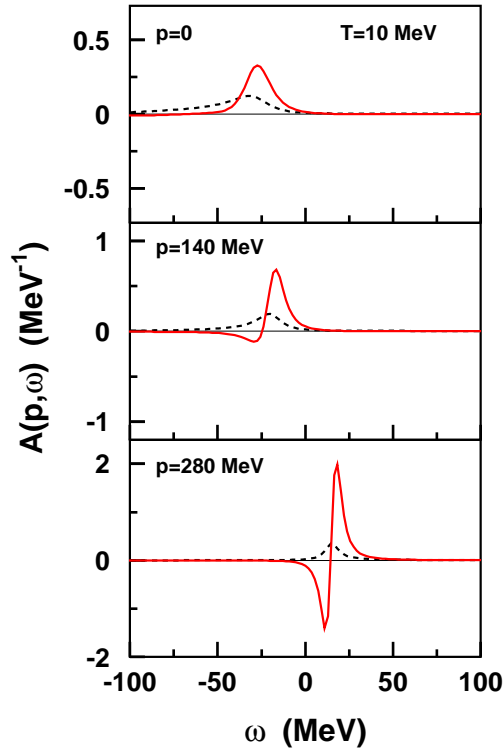}
\caption{(Color online)
The spectral function $A(\mathbf{p},\omega)$ (dashed line) and the 
function $B(\mathbf{p},\omega)$ (solid line)
 calculated for symmetric nuclear matter at $T=10$ MeV.}
\label{fig:spectral}
\end{center}
\end{figure}
For the weight function $B(\mathbf{p},\omega)$ a similar separation is 
performed
\begin{equation}
B(\mathbf{p},\omega)=B_{bg}(\mathbf{p},\omega)+2 \pi \, W_p \,
\mbox{Re}\Sigma(\mathbf{p},\omega) \, \delta^{'}(\omega-\omega_p)
+\frac{ W_p \,
\mbox{Im}\Sigma(\mathbf{p},\omega)}{(\omega-\omega_p)^2} \ ,
\end{equation}
which conserves the property (\ref{eq:intb}).

\begin{table}[htbp] 
\begin{center} 
\begin{tabular}{|c||c|c|c|c|c|c|c|} 
\hline $\; T   \;$  & $\; E_{GK}/N \;$ & $\;  E_{diag}/N \;$ &
$\; P_I/ \rho  \;$ & 
$\; P_{II}/ \rho \;$ & 
$\; P_{tot}/ \rho  \;$ &
$\; P_{quasi}/ \rho \; $&  $\; P_{BF}/\rho \; $
\\ \hline 
\hline $0$ & $-15.80$ & $-16.63$ &  $-40.19$ & $32.50$  & $-7.69$ & $-0.85$
& $-5.58$ 
\\ 
\hline $2$ & $-15.15$ & $-16.29$ & $-38.40$ & $32.54$  & $-5.86$ & $-0.78$
& $-5.4$
\\ 
\hline $5$ & $-14.40$ & $-15.24$ &  $-37.83$ & $32.35$  & $-5.48$ & $-0.74$
& $-5.2$
\\ 
\hline $10$ & $-11.15$ & $-11.72$ & $-34.02$ & $31.36$  & $-2.66$ & $-0.49$
& $-3.35$
\\
\hline $20$ & $-1.29$ & $-1.21$ & $-24.43$ & $30.92$  & $\:\:6.49$ & $\:\:0.17$
& $4.74$
\\
\hline
\end{tabular} 
\end{center} 
\caption{Results (at $\rho = \rho_0$) of the internal energy and the  pressure 
 at zero and finite temperature
(all values in MeVs) with the CD-Bonn interaction. 
The second and third colum represent the internal energy
 per particle obtained from the Galitskii-Koltun's sum rule (\ref{gksumrule})
and from the diagram summation (\ref{eq:be}).
The other  columns correspond to the partial results 
(\ref{eq:p1}) and (\ref{eq:p2}), to the sum (\ref{eq:pre}) and to the
quasiparticle expression (\ref{eq:QPpress}) for the pressure.
In the  last column are quoted 
the results of Baldo and Ferreira \cite{Baldo:Ferreira}, obtained with
the Argonne $v_{14}$ interaction.}
\label{tab:pressure} 
\end{table}

The diagrams contributing to the $\Phi$ functional are summed in the following 
way. One notes that for the $T-$matrix approximation 
 the expressions for the interaction energy and the 
functional  $\Phi$ differ by the factor $1/n$ where $n$ is 
the number of interaction lines in the diagram (Figs. \ref{fig:tphi} and
\ref{fig:diagv}). In that case, the functional $\Phi$ can be 
obtained from the formula for the interaction energy
\begin{equation}
\label{eq:ourphi}
\Phi=\int_0^1
\frac{d\lambda}{\lambda} <H_{pot}(\lambda V, G_{\lambda=1})> \ \ .
\end{equation}
In the above formula  the interaction potential is multiplied by the 
factor $\lambda$ but the
propagator $G$ is the dressed nucleon propagator corresponding to the
system with the full strength of interactions $(\lambda=1)$. The 
method for calculating the functional $\Phi$ that we use (\ref{eq:ourphi})
should not be confused with the textbook expression for the pressure 
of an interacting system \cite{baym,fw}
\begin{equation}
PV=P_0V+\int_0^1
\frac{d\lambda}{\lambda} <H_{pot}(\lambda V, G_{\lambda})> \ \ ,
\end{equation}
 where $P_0$ is the pressure of a noninteracting system and the
average interaction $<H_{pot}>$
 is calculated in a system with the  potential reduced 
by a factor $\lambda$ and using the propagator $G_\lambda$ calculated 
self-consistently in the system with reduced interactions.

The results of the calculations, for different temperatures, 
are shown in table \ref{tab:pressure}. % and in figure \ref{fig:pressure}.
We observe a steady  increase of the pressure with the temperature. Only at 
the lowest temperatures some modifications of this behavior are visible,
which may be due to changes in the single-particle properties close to the
critical point for the pairing transition
\cite{Alm:1996ps,Schnell:1999tu,Bozek:1999rv,Bozek:2002jw,
Guttormsen:2002nz,Dean:2002zx}.
 We  compare our results with the pressure of a gas of quasiparticles in
a mean-field potential
\begin{equation}
\label{eq:QPpress}
P_{qp}=\int \frac{d^3p}{(2\pi)^3} f(\omega_{p}-\mu) 
\left [
\frac{{p}}{3} \frac{d \omega_{p}}{dp}
+\frac{1}{2} \Sigma_p \right ] \: ,
\end{equation}
with $\Sigma_p = \Sigma(p,\omega_p)=\omega_{p} - p^2/2m + \mu $ .
The expression (\ref{eq:QPpress}) is the formula for the pressure in the 
mean-field approximation, with $\Sigma_p$ taken for the mean-field. 
The results are very different from the full calculation 
with dispersive self-energies 
and spectral functions. Hence we conclude that, using the CD-Bonn
potential, the pressure cannot be obtained 
from  the mean-field-like formula (\ref{eq:QPpress}) around saturation 
density.  
At low densities or high temperatures the pressure in a interacting
nucleon gas can be obtained in a model independent way from the 
elastic scattering phase
shifts \cite{bu,prattsu,Horowitz:2005nd}. However at the
saturation density (at low temperatures) the pressure cannot be 
reliably calculated from a viral
expansion. The  pressure in hot nuclear matter has been obtained using
two-body Argonne $v_{14}$ interaction in the Bloch-De Dominicis approach 
\cite{Baldo:Ferreira}.
The results are qualitatively similar, with a negative
value of the pressure at $T=0$ and $\rho\simeq 0.16$ fm$^{-3}$. This shows
the need to include three-body forces for a reliable description of the
thermodynamics of the nuclear matter.

%\begin{figure}[h]
%\begin{center}
%\includegraphics[width=8cm]{totpre.ps}
%\caption{$P_{tot}/ \rho $ as a function of temperature.}
%\label{fig:pressure}
%\end{center}
%\end{figure}

\begin{figure}[h]
\begin{center}
\includegraphics[width=8cm]{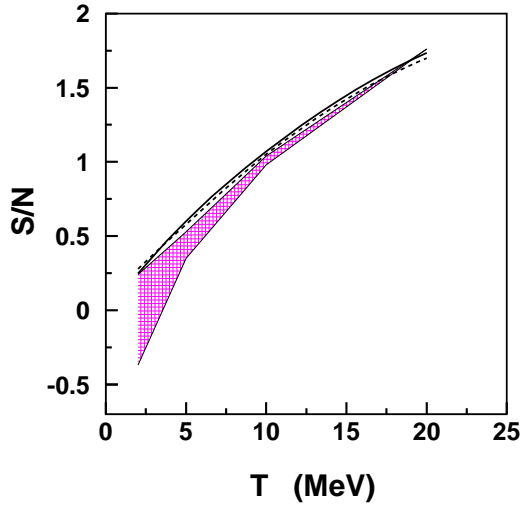}
\caption{(Color online) Entropy per baryon 
 as a function of the temperature. The hatched band denotes our 
estimate of the entropy  corresponding to the two values $S_{GK}$ and
$S_{diag}$ from Table \ref{tab:entropy}. The solid line denotes the result 
for the free Fermi gas with the in-medium effective mass and the dashed line
represents the result of \cite{barcelona}  (Eq. \ref{eq:ssimp}).}
\label{fig:entropy}
\end{center}
\end{figure}

We compute the entropy through the thermodynamic relation
\begin{equation}
\frac{S}{N} = \frac{1}{T} 
\left [
\frac{E}{N} + \frac{P_{tot}}{\rho} - \mu
\right ] \: .
\end{equation}
The results are displayed in table \ref{tab:entropy}.
\begin{table}[htbp] 
\begin{center} 
\begin{tabular}{|c||*{6}{c|}} 
\hline $\; T \: \: (\mbox{MeV}) \;$  &
$\; \; S_{GK}/ N \; \;$ &
$\; \; S_{diag}/ N \; \;$ &
$\; \; S_{free}/ N \; \;$ &
$m^{\star} \: \: (\mbox{MeV})$ &
$\; \; S_{free^{\star}}/ N \; \;$ &
$\; \; S_{DQ}/ N \; \;$
\\ \hline 
\hline $2$ & $0.24$ & $-0.37$ & $0.27$ & $873$ & $0.24$ & $0.28$
\\ 
\hline $5$ & $0.53$ & $0.35$ & $0.66$ & $853$ & $0.60$ & $0.58$
\\ 
\hline $10$ & $1.04$ & $0.98$ & $1.22$ & $802$ & $1.07$ & $1.05$
\\ 
\hline $20$ & $1.76$ & $1.76$ & $2.02$ & $745$ & $1.74$ & $1.70$
\\
\hline
\end{tabular} 
\end{center} 
\caption{Entropy per baryon. The results for 
 the interacting system in the $T$-matrix
  approximation using the Galitskii-Koltun's sum rule 
  (\ref{gksumrule}) and Eq. (\ref{eq:be}) expressions for 
for the internal energy are shown in 
 columns $S_{GK}/N$ and $S_{diag}/N$ respectively.
$S_{DQ}/N$ denotes the results of \cite{barcelona}  (Eq. \ref{eq:ssimp}).
$S_{free}/N$ and $S_{free}^\star/N$ are the entropies per baryon in a 
free Fermi gas with the free and in medium masses respectively.}
\label{tab:entropy} 
\end{table}
The error in the calculation  of $TS$ at each temperature can be estimated by
comparing the results obtained with the two expressions for 
the internal energy 
(\ref{eq:be}) and (\ref{gksumrule}). The difference is of the order of
$1$ MeV, also at zero temperature we find $|TS|\simeq 1\,\mbox{MeV}\neq 0$.
The entropy can be estimated reliably only for $T\ge 5$~MeV, with the
uncertainty shown as the hatched band in Fig. \ref{fig:entropy}.

We compare these results with other methods of calculating the entropy:
\begin{enumerate}
\item The dynamical quasiparticle formula \cite{pethickentropy}
\begin{equation}
\label{eq:ssimp}
\frac{S_{DQ}}{N} =\frac{1}{\rho}
 \int \frac{d^3p}{(2 \pi)^3} \frac{d \omega}{2 \pi}
\sigma(\omega) \left[A({\bf
 p},\omega)\left(1-\frac{\partial\mbox{Re}\Sigma^{R}
(\mathbf{p},\omega)}
{\partial\omega}\right) + \frac{\partial\mbox{Re}G^{R}
(\mathbf{p},\omega)}
{\partial\omega}\Gamma(\mathbf{p},\omega)\right] \ \ ,
\end{equation}
where
\begin{equation}
\sigma(\omega) = - f(\omega) \ln [f(\omega)] 
- [1-f(\omega)] \ln [1-f(\omega)] \: .
\end{equation}
It has been shown that this one-body formula gives results close to the 
complete expression for the entropy \cite{barcelona}.
\item The entropy for a free Fermi gas
\begin{equation}
\frac{S_{free}}{N} = \frac{1}{\rho} \int \frac{d^3p}{(2\pi)^3} \ 
 \sigma\left(\frac{p^2}{2m}\right) \ \ .
\end{equation}
\item The entropy calculated as for a free Fermi gas but using the effective
mass $m^*$ instead of the rest mass $m$
\begin{equation}
\frac{S_{{free}^\star}}{N} =\frac{1}{\rho} \int \frac{d^3p}{(2\pi)^3}\
\sigma\left(\frac{p^2}{2m^\star}\right)   \ \ .
\end{equation}
The effective mass $m^*$ is determined at each temperature by
\begin{equation}
{\left(
\frac{\partial {\omega}_p}{\partial p^2}
\right)}_{p=p_F}
= \frac{1}{2 m^\star} \: .
\end{equation}
\end{enumerate}
 Remarkably,  we find that    the expression for 
the entropy of the free Fermi gas is very similar to the result of the full
calculation for the interacting system, if the
change of  the effective mass in the system is taken into account.
The last observation  simplifies  significantly 
      the modeling of the evolution of protoneutron stars 
\cite{ Prakash:1996xs}, since relations between the entropy
per baryon and the temperature derived in the case of a fermion gas 
can be used in hot nuclear matter. As observed in \cite{barcelona} the 
quasiparticle expression (\ref{eq:ssimp}) for the entropy 
 \cite{pethickentropy} follows closely the full
result.

\section{Discussion}

We study the properties of nuclear matter with short range correlations
at finite temperatures up to $T=20$~MeV. We calculate  the internal energy,
the pressure, and the entropy. 
It is to our knowledge the first calculation of the pressure and
of the entropy in the thermodynamically consistent $T-$matrix approximation
in  nuclear matter. Besides the usually employed Galitskii-Koltun's sum
rule for the internal energy,
 we  perform a summation of diagrams corresponding to  the expectation value of
the interaction energy. The two methods yield similar results, up to 
a difference
of about $1$ MeV which can be attributed to numerical inaccuracies and to the
angular averaging used in the partial wave expansion of the in medium
$T-$matrix. The calculation of the pressure requires a summation of a
different set of diagrams, due to different numerical factors. The result is
most easily obtained by an integration over an artificial parameter $\lambda$
multiplying the interaction lines, while keeping the propagators dressed as in
the fully correlated system. From the pressure and the internal energy we
obtain the entropy per baryon at temperatures $T\ge 5$ MeV with an
uncertainty that we estimate to be $1\,\mbox{MeV}/T$. 
 The entropy of the free Fermi gas
turns out to be close to the result of the full calculation  if the 
change of the effective mass in the medium is taken into account.

\bibliography{mojbib}

\begin{thebibliography}{41}
\expandafter\ifx\csname natexlab\endcsname\relax\def\natexlab#1{#1}\fi
\expandafter\ifx\csname bibnamefont\endcsname\relax
  \def\bibnamefont#1{#1}\fi
\expandafter\ifx\csname bibfnamefont\endcsname\relax
  \def\bibfnamefont#1{#1}\fi
\expandafter\ifx\csname citenamefont\endcsname\relax
  \def\citenamefont#1{#1}\fi
\expandafter\ifx\csname url\endcsname\relax
  \def\url#1{\texttt{#1}}\fi
\expandafter\ifx\csname urlprefix\endcsname\relax\def\urlprefix{URL }\fi
\providecommand{\bibinfo}[2]{#2}
\providecommand{\eprint}[2][]{\url{#2}}

\bibitem[{\citenamefont{Baym and Kadanoff}(1961)}]{kadBaym}
\bibinfo{author}{\bibfnamefont{G.}~\bibnamefont{Baym}} \bibnamefont{and}
  \bibinfo{author}{\bibfnamefont{L.}~\bibnamefont{Kadanoff}},
  \bibinfo{journal}{Phys. Rev.} \textbf{\bibinfo{volume}{124}},
  \bibinfo{pages}{287} (\bibinfo{year}{1961}).

\bibitem[{\citenamefont{Baym}(1962)}]{baym}
\bibinfo{author}{\bibfnamefont{G.}~\bibnamefont{Baym}}, \bibinfo{journal}{Phys.
  Rev.} \textbf{\bibinfo{volume}{127}}, \bibinfo{pages}{1392}
  (\bibinfo{year}{1962}).

\bibitem[{\citenamefont{Bo\.zek and Czerski}(2001)}]{Bozek:2001tz}
\bibinfo{author}{\bibfnamefont{P.}~\bibnamefont{Bo\.zek}} \bibnamefont{and}
  \bibinfo{author}{\bibfnamefont{P.}~\bibnamefont{Czerski}},
  \bibinfo{journal}{Eur. Phys. J.} \textbf{\bibinfo{volume}{A11}},
  \bibinfo{pages}{271} (\bibinfo{year}{2001}),
  \eprint[http://arXiv.org/abs]{nucl-th/0102020}.

\bibitem[{\citenamefont{Bo\.zek}(2002{\natexlab{a}})}]{Bozek:2002mw}
\bibinfo{author}{\bibfnamefont{P.}~\bibnamefont{Bo\.zek}},
  \bibinfo{journal}{Eur. Phys. J.} \textbf{\bibinfo{volume}{A15}},
  \bibinfo{pages}{325} (\bibinfo{year}{2002}{\natexlab{a}}),
  \eprint{nucl-th/0204034}.

\bibitem[{\citenamefont{Hugenholz and Hove}(1958)}]{hvh}
\bibinfo{author}{\bibfnamefont{N.}~\bibnamefont{Hugenholz}} \bibnamefont{and}
  \bibinfo{author}{\bibfnamefont{L.~V.} \bibnamefont{Hove}},
  \bibinfo{journal}{Physica} \textbf{\bibinfo{volume}{24}},
  \bibinfo{pages}{363} (\bibinfo{year}{1958}).

\bibitem[{\citenamefont{Luttinger}(1960)}]{luttinger2}
\bibinfo{author}{\bibfnamefont{J.~M.} \bibnamefont{Luttinger}},
  \bibinfo{journal}{Phys. Rev.} \textbf{\bibinfo{volume}{119}},
  \bibinfo{pages}{1151} (\bibinfo{year}{1960}).

\bibitem[{\citenamefont{Bo\.zek and Czerski}(2003)}]{Bozek:2002tz}
\bibinfo{author}{\bibfnamefont{P.}~\bibnamefont{Bo\.zek}} \bibnamefont{and}
  \bibinfo{author}{\bibfnamefont{P.}~\bibnamefont{Czerski}},
  \bibinfo{journal}{Acta Phys. Polon.} \textbf{\bibinfo{volume}{B34}},
  \bibinfo{pages}{2759} (\bibinfo{year}{2003}), \eprint{nucl-th/0212035}.

\bibitem[{\citenamefont{Bo\.zek and Czerski}(2002)}]{Bozek:2002ry}
\bibinfo{author}{\bibfnamefont{P.}~\bibnamefont{Bo\.zek}} \bibnamefont{and}
  \bibinfo{author}{\bibfnamefont{P.}~\bibnamefont{Czerski}},
  \bibinfo{journal}{Phys. Rev.} \textbf{\bibinfo{volume}{C66}},
  \bibinfo{pages}{027301} (\bibinfo{year}{2002}), \eprint{nucl-th/0204012}.

\bibitem[{\citenamefont{Dewulf et~al.}(2002)\citenamefont{Dewulf, Van~Neck, and
  Waroquier}}]{Dewulf:2002gi}
\bibinfo{author}{\bibfnamefont{Y.}~\bibnamefont{Dewulf}},
  \bibinfo{author}{\bibfnamefont{D.}~\bibnamefont{Van~Neck}}, \bibnamefont{and}
  \bibinfo{author}{\bibfnamefont{M.}~\bibnamefont{Waroquier}},
  \bibinfo{journal}{Phys. Rev.} \textbf{\bibinfo{volume}{C65}},
  \bibinfo{pages}{054316} (\bibinfo{year}{2002}).

\bibitem[{\citenamefont{Dewulf et~al.}(2003)\citenamefont{Dewulf, Dickhoff,
  Van~Neck, Stoddard, and Waroquier}}]{Dewulf:2003nj}
\bibinfo{author}{\bibfnamefont{Y.}~\bibnamefont{Dewulf}},
  \bibinfo{author}{\bibfnamefont{W.~H.} \bibnamefont{Dickhoff}},
  \bibinfo{author}{\bibfnamefont{D.}~\bibnamefont{Van~Neck}},
  \bibinfo{author}{\bibfnamefont{E.~R.} \bibnamefont{Stoddard}},
  \bibnamefont{and}
  \bibinfo{author}{\bibfnamefont{M.}~\bibnamefont{Waroquier}},
  \bibinfo{journal}{Phys. Rev. Lett.} \textbf{\bibinfo{volume}{90}},
  \bibinfo{pages}{152501} (\bibinfo{year}{2003}), \eprint{nucl-th/0303047}.

\bibitem[{\citenamefont{Akmal et~al.}(1998)\citenamefont{Akmal, Pandharipande,
  and Ravenhall}}]{vcs2}
\bibinfo{author}{\bibfnamefont{A.}~\bibnamefont{Akmal}},
  \bibinfo{author}{\bibfnamefont{V.~R.} \bibnamefont{Pandharipande}},
  \bibnamefont{and} \bibinfo{author}{\bibfnamefont{D.~G.}
  \bibnamefont{Ravenhall}}, \bibinfo{journal}{Phys. Rev. C}
  \textbf{\bibinfo{volume}{58}}, \bibinfo{pages}{1804} (\bibinfo{year}{1998}).

\bibitem[{\citenamefont{Wiringa et~al.}(1988)\citenamefont{Wiringa, Fiks, and
  Fabrocini}}]{vcs1}
\bibinfo{author}{\bibfnamefont{R.~B.} \bibnamefont{Wiringa}},
  \bibinfo{author}{\bibfnamefont{V.}~\bibnamefont{Fiks}}, \bibnamefont{and}
  \bibinfo{author}{\bibfnamefont{A.}~\bibnamefont{Fabrocini}},
  \bibinfo{journal}{Phys. Rev. C} \textbf{\bibinfo{volume}{38}},
  \bibinfo{pages}{1010} (\bibinfo{year}{1988}).

\bibitem[{\citenamefont{Baldo and Burgio}(2001)}]{baldonstar}
\bibinfo{author}{\bibfnamefont{M.}~\bibnamefont{Baldo}} \bibnamefont{and}
  \bibinfo{author}{\bibfnamefont{G.~F.} \bibnamefont{Burgio}}, in
  \emph{\bibinfo{booktitle}{Microscopic Theory of Nuclear Equation of State and
  Neutron Star Structure}}, edited by
  \bibinfo{editor}{\bibfnamefont{D.}~\bibnamefont{Blaschke}},
  \bibinfo{editor}{\bibfnamefont{N.}~\bibnamefont{Glendenning}},
  \bibnamefont{and} \bibinfo{editor}{\bibfnamefont{A.}~\bibnamefont{Sedrakian}}
  (\bibinfo{publisher}{Springer}, \bibinfo{address}{Heidelberg},
  \bibinfo{year}{2001}), vol. \bibinfo{volume}{578} of
  \emph{\bibinfo{series}{Lecture Notes in Physics}},
  \eprint[http://arXiv.org/abs]{nucl-th/0012014}.

\bibitem[{\citenamefont{Baldo et~al.}(2002)\citenamefont{Baldo, Fiasconaro,
  Song, Giansiracusa, and Lombardo}}]{Baldo:2001mw}
\bibinfo{author}{\bibfnamefont{M.}~\bibnamefont{Baldo}},
  \bibinfo{author}{\bibfnamefont{A.}~\bibnamefont{Fiasconaro}},
  \bibinfo{author}{\bibfnamefont{H.~Q.} \bibnamefont{Song}},
  \bibinfo{author}{\bibfnamefont{G.}~\bibnamefont{Giansiracusa}},
  \bibnamefont{and} \bibinfo{author}{\bibfnamefont{U.}~\bibnamefont{Lombardo}},
  \bibinfo{journal}{Phys. Rev.} \textbf{\bibinfo{volume}{C65}},
  \bibinfo{pages}{017303} (\bibinfo{year}{2002}).

\bibitem[{\citenamefont{Baldo and Ferreira}(1998)}]{Baldo:Ferreira}
\bibinfo{author}{\bibfnamefont{M.}~\bibnamefont{Baldo}} \bibnamefont{and}
  \bibinfo{author}{\bibfnamefont{L.~S.} \bibnamefont{Ferreira}},
  \bibinfo{journal}{Phys. Rev.} \textbf{\bibinfo{volume}{C59}},
  \bibinfo{pages}{682} (\bibinfo{year}{1998}).

\bibitem[{\citenamefont{Zuo et~al.}(2006)\citenamefont{Zuo, Li, Lombardo, Lu,
  and Schulze}}]{Zuo:2006wx}
\bibinfo{author}{\bibfnamefont{W.}~\bibnamefont{Zuo}},
  \bibinfo{author}{\bibfnamefont{Z.~H.} \bibnamefont{Li}},
  \bibinfo{author}{\bibfnamefont{U.}~\bibnamefont{Lombardo}},
  \bibinfo{author}{\bibfnamefont{G.~C.} \bibnamefont{Lu}}, \bibnamefont{and}
  \bibinfo{author}{\bibfnamefont{H.~J.} \bibnamefont{Schulze}},
  \bibinfo{journal}{Phys. Rev.} \textbf{\bibinfo{volume}{C73}},
  \bibinfo{pages}{035208} (\bibinfo{year}{2006}).

\bibitem[{\citenamefont{Bo\.zek}(1999{\natexlab{a}})}]{Bozek:1998su}
\bibinfo{author}{\bibfnamefont{P.}~\bibnamefont{Bo\.zek}},
  \bibinfo{journal}{Phys. Rev.} \textbf{\bibinfo{volume}{C59}},
  \bibinfo{pages}{2619} (\bibinfo{year}{1999}{\natexlab{a}}),
  \eprint[http://arXiv.org/abs]{nucl-th/9811073}.

\bibitem[{\citenamefont{Bo\.zek}(2002{\natexlab{b}})}]{Bozek:2002em}
\bibinfo{author}{\bibfnamefont{P.}~\bibnamefont{Bo\.zek}},
  \bibinfo{journal}{Phys. Rev.} \textbf{\bibinfo{volume}{C65}},
  \bibinfo{pages}{054306} (\bibinfo{year}{2002}{\natexlab{b}}),
  \eprint[http://arXiv.org/abs]{nucl-th/0201086}.

\bibitem[{\citenamefont{Frick and {M\"uther}}(2003)}]{Frick:2003sd}
\bibinfo{author}{\bibfnamefont{T.}~\bibnamefont{Frick}} \bibnamefont{and}
  \bibinfo{author}{\bibfnamefont{H.}~\bibnamefont{{M\"uther}}},
  \bibinfo{journal}{Phys. Rev.} \textbf{\bibinfo{volume}{C68}},
  \bibinfo{pages}{034310} (\bibinfo{year}{2003}), \eprint{nucl-th/0306009}.

\bibitem[{\citenamefont{Frick et~al.}(2004)\citenamefont{Frick, {M\"uther}, and
  Polls}}]{Frick2}
\bibinfo{author}{\bibfnamefont{T.}~\bibnamefont{Frick}},
  \bibinfo{author}{\bibfnamefont{H.}~\bibnamefont{{M\"uther}}},
  \bibnamefont{and} \bibinfo{author}{\bibfnamefont{A.}~\bibnamefont{Polls}}
  (\bibinfo{year}{2004}), \eprint{nucl-th/0401015}.

\bibitem[{\citenamefont{Frick et~al.}(2005)\citenamefont{Frick, {M\"uther},
  Rios, Polls, and Ramos}}]{Frick:2004th}
\bibinfo{author}{\bibfnamefont{T.}~\bibnamefont{Frick}},
  \bibinfo{author}{\bibfnamefont{H.}~\bibnamefont{{M\"uther}}},
  \bibinfo{author}{\bibfnamefont{A.}~\bibnamefont{Rios}},
  \bibinfo{author}{\bibfnamefont{A.}~\bibnamefont{Polls}}, \bibnamefont{and}
  \bibinfo{author}{\bibfnamefont{A.}~\bibnamefont{Ramos}},
  \bibinfo{journal}{Phys. Rev.} \textbf{\bibinfo{volume}{C71}},
  \bibinfo{pages}{014313} (\bibinfo{year}{2005}), \eprint{nucl-th/0409067}.

\bibitem[{\citenamefont{Keldysh}(1964)}]{keldysh}
\bibinfo{author}{\bibfnamefont{L.~V.} \bibnamefont{Keldysh}},
  \bibinfo{journal}{Zh. Eksp. Teor. Fiz.} \textbf{\bibinfo{volume}{47}},
  \bibinfo{pages}{1515} (\bibinfo{year}{1964}).

\bibitem[{\citenamefont{Danielewicz}(1984)}]{Danielewicz:1982kk}
\bibinfo{author}{\bibfnamefont{P.}~\bibnamefont{Danielewicz}},
  \bibinfo{journal}{Annals Phys.} \textbf{\bibinfo{volume}{152}},
  \bibinfo{pages}{239} (\bibinfo{year}{1984}).

\bibitem[{\citenamefont{Galitskii and Migdal}(1958)}]{galitskii}
\bibinfo{author}{\bibfnamefont{V.~M.} \bibnamefont{Galitskii}}
  \bibnamefont{and} \bibinfo{author}{\bibfnamefont{A.~B.}
  \bibnamefont{Migdal}}, \bibinfo{journal}{Zh. Eksp. Teor. Fiz.}
  \textbf{\bibinfo{volume}{34}}, \bibinfo{pages}{139} (\bibinfo{year}{1958}).

\bibitem[{\citenamefont{Martin and Schwinger}(1959)}]{martinschwinger}
\bibinfo{author}{\bibfnamefont{P.~C.} \bibnamefont{Martin}} \bibnamefont{and}
  \bibinfo{author}{\bibfnamefont{J.}~\bibnamefont{Schwinger}},
  \bibinfo{journal}{Phys. Rev.} \textbf{\bibinfo{volume}{115}},
  \bibinfo{pages}{1342} (\bibinfo{year}{1959}).

\bibitem[{\citenamefont{Koltun}(1972)}]{koltun}
\bibinfo{author}{\bibfnamefont{D.~S.} \bibnamefont{Koltun}},
  \bibinfo{journal}{Phys. Rev. Lett.} \textbf{\bibinfo{volume}{28}},
  \bibinfo{pages}{182} (\bibinfo{year}{1972}).

\bibitem[{\citenamefont{Schiller et~al.}(1999)\citenamefont{Schiller, Muther,
  and Czerski}}]{Schiller:1998ff}
\bibinfo{author}{\bibfnamefont{E.}~\bibnamefont{Schiller}},
  \bibinfo{author}{\bibfnamefont{H.}~\bibnamefont{Muther}}, \bibnamefont{and}
  \bibinfo{author}{\bibfnamefont{P.}~\bibnamefont{Czerski}},
  \bibinfo{journal}{Phys. Rev.} \textbf{\bibinfo{volume}{C59}},
  \bibinfo{pages}{2934} (\bibinfo{year}{1999}), \eprint{nucl-th/9812011}.

\bibitem[{\citenamefont{Fetter and Walecka}(1971)}]{fw}
\bibinfo{author}{\bibfnamefont{A.~L.} \bibnamefont{Fetter}} \bibnamefont{and}
  \bibinfo{author}{\bibfnamefont{J.~D.} \bibnamefont{Walecka}},
  \emph{\bibinfo{title}{Quantum theory of many particle system}}
  (\bibinfo{publisher}{McGraw-Hill}, \bibinfo{address}{New York},
  \bibinfo{year}{1971}).

\bibitem[{\citenamefont{Weinhold et~al.}(1998)\citenamefont{Weinhold, Friman,
  and Norenberg}}]{Weinhold:1997ig}
\bibinfo{author}{\bibfnamefont{W.}~\bibnamefont{Weinhold}},
  \bibinfo{author}{\bibfnamefont{B.}~\bibnamefont{Friman}}, \bibnamefont{and}
  \bibinfo{author}{\bibfnamefont{W.}~\bibnamefont{Norenberg}},
  \bibinfo{journal}{Phys. Lett.} \textbf{\bibinfo{volume}{B433}},
  \bibinfo{pages}{236} (\bibinfo{year}{1998}), \eprint{nucl-th/9710014}.

\bibitem[{\citenamefont{Alm et~al.}(1996)\citenamefont{Alm, {R\"opke}, Schnell,
  Kwong, and Kohler}}]{Alm:1996ps}
\bibinfo{author}{\bibfnamefont{T.}~\bibnamefont{Alm}},
  \bibinfo{author}{\bibfnamefont{G.}~\bibnamefont{{R\"opke}}},
  \bibinfo{author}{\bibfnamefont{A.}~\bibnamefont{Schnell}},
  \bibinfo{author}{\bibfnamefont{N.~H.} \bibnamefont{Kwong}}, \bibnamefont{and}
  \bibinfo{author}{\bibfnamefont{H.~S.} \bibnamefont{Kohler}},
  \bibinfo{journal}{Phys. Rev.} \textbf{\bibinfo{volume}{C53}},
  \bibinfo{pages}{2181} (\bibinfo{year}{1996}),
  \eprint[http://arXiv.org/abs]{nucl-th/9511039}.

\bibitem[{\citenamefont{Schnell et~al.}(1999)\citenamefont{Schnell, {R\"opke},
  and Schuck}}]{Schnell:1999tu}
\bibinfo{author}{\bibfnamefont{A.}~\bibnamefont{Schnell}},
  \bibinfo{author}{\bibfnamefont{G.}~\bibnamefont{{R\"opke}}},
  \bibnamefont{and} \bibinfo{author}{\bibfnamefont{P.}~\bibnamefont{Schuck}},
  \bibinfo{journal}{Phys. Rev. Lett.} \textbf{\bibinfo{volume}{83}},
  \bibinfo{pages}{1926} (\bibinfo{year}{1999}),
  \eprint[http://arXiv.org/abs]{nucl-th/9902038}.

\bibitem[{\citenamefont{Bo\.zek}(1999{\natexlab{b}})}]{Bozek:1999rv}
\bibinfo{author}{\bibfnamefont{P.}~\bibnamefont{Bo\.zek}},
  \bibinfo{journal}{Nucl. Phys.} \textbf{\bibinfo{volume}{A657}},
  \bibinfo{pages}{187} (\bibinfo{year}{1999}{\natexlab{b}}),
  \eprint[http://arXiv.org/abs]{nucl-th/9902019}.

\bibitem[{\citenamefont{Bo\.zek}(2002{\natexlab{c}})}]{Bozek:2002jw}
\bibinfo{author}{\bibfnamefont{P.}~\bibnamefont{Bo\.zek}},
  \bibinfo{journal}{Phys. Lett.} \textbf{\bibinfo{volume}{B551}},
  \bibinfo{pages}{93} (\bibinfo{year}{2002}{\natexlab{c}}),
  \eprint[http://arXiv.org/abs]{nucl-th/0202045}.

\bibitem[{\citenamefont{Guttormsen et~al.}(2003)}]{Guttormsen:2002nz}
\bibinfo{author}{\bibfnamefont{M.}~\bibnamefont{Guttormsen}}
  \bibnamefont{et~al.}, \bibinfo{journal}{Phys. Rev.}
  \textbf{\bibinfo{volume}{C68}}, \bibinfo{pages}{034311}
  (\bibinfo{year}{2003}), \eprint{nucl-ex/0209013}.

\bibitem[{\citenamefont{Dean and Hjorth-Jensen}(2003)}]{Dean:2002zx}
\bibinfo{author}{\bibfnamefont{D.~J.} \bibnamefont{Dean}} \bibnamefont{and}
  \bibinfo{author}{\bibfnamefont{M.}~\bibnamefont{Hjorth-Jensen}},
  \bibinfo{journal}{Rev. Mod. Phys.} \textbf{\bibinfo{volume}{75}},
  \bibinfo{pages}{607} (\bibinfo{year}{2003}), \eprint{nucl-th/0210033}.

\bibitem[{\citenamefont{Beth and Uhlenbeck}(1937)}]{bu}
\bibinfo{author}{\bibfnamefont{U.}~\bibnamefont{Beth}} \bibnamefont{and}
  \bibinfo{author}{\bibfnamefont{G.~E.} \bibnamefont{Uhlenbeck}},
  \bibinfo{journal}{Physica} \textbf{\bibinfo{volume}{4}}, \bibinfo{pages}{915}
  (\bibinfo{year}{1937}).

\bibitem[{\citenamefont{Horowitz and Schwenk}(2005)}]{Horowitz:2005nd}
\bibinfo{author}{\bibfnamefont{C.~J.} \bibnamefont{Horowitz}} \bibnamefont{and}
  \bibinfo{author}{\bibfnamefont{A.}~\bibnamefont{Schwenk}}
  (\bibinfo{year}{2005}), \eprint{nucl-th/0507033}.

\bibitem[{\citenamefont{Pratt et~al.}(1987)\citenamefont{Pratt, Siemens, and
  Usmani}}]{prattsu}
\bibinfo{author}{\bibfnamefont{S.}~\bibnamefont{Pratt}},
  \bibinfo{author}{\bibfnamefont{P.}~\bibnamefont{Siemens}}, \bibnamefont{and}
  \bibinfo{author}{\bibfnamefont{Q.~N.} \bibnamefont{Usmani}},
  \bibinfo{journal}{Phys. Lett. B} \textbf{\bibinfo{volume}{189}},
  \bibinfo{pages}{1} (\bibinfo{year}{1987}).

\bibitem[{\citenamefont{Rios et~al.}(2006)\citenamefont{Rios, Polls, Ramos, and
  M{\"u}ther}}]{barcelona}
\bibinfo{author}{\bibfnamefont{A.}~\bibnamefont{Rios}},
  \bibinfo{author}{\bibfnamefont{A.}~\bibnamefont{Polls}},
  \bibinfo{author}{\bibfnamefont{A.}~\bibnamefont{Ramos}}, \bibnamefont{and}
  \bibinfo{author}{\bibfnamefont{H.}~\bibnamefont{M{\"u}ther}}
  (\bibinfo{year}{2006}), \eprint{nucl-th/0605080}.

\bibitem[{\citenamefont{Carneiro and Pethick}(1975)}]{pethickentropy}
\bibinfo{author}{\bibfnamefont{G.}~\bibnamefont{Carneiro}} \bibnamefont{and}
  \bibinfo{author}{\bibfnamefont{C.}~\bibnamefont{Pethick}},
  \bibinfo{journal}{Phys. Rev} \textbf{\bibinfo{volume}{D11}},
  \bibinfo{pages}{1106} (\bibinfo{year}{1975}).

\bibitem[{\citenamefont{Prakash et~al.}(1997)}]{Prakash:1996xs}
\bibinfo{author}{\bibfnamefont{M.}~\bibnamefont{Prakash}} \bibnamefont{et~al.},
  \bibinfo{journal}{Phys. Rept.} \textbf{\bibinfo{volume}{280}},
  \bibinfo{pages}{1} (\bibinfo{year}{1997}), \eprint{nucl-th/9603042}.

\end{thebibliography}

\end{document}